\documentclass[prl,superscriptaddress,showpacs,twocolumn]{revtex4}
\usepackage{graphicx}
\usepackage{amsmath}
\usepackage{color}

\begin{document}

\title{Non-local Coulomb interactions and metal-insulator transition in Ti$_2$O$_3$: \\
a cluster LDA+DMFT approach.}
\author{A.I. Poteryaev}
\affiliation{ESM, University of Nijmegen, NL-6525 ED Nijmegen, The Netherlands}
\author{A.I. Lichtenstein}
\affiliation{ESM, University of Nijmegen, NL-6525 ED Nijmegen, The Netherlands}
\author{G. Kotliar}
\affiliation{Serin Physics Laboratory, Rutgers University, Piscataway, New Jersey 08855}
\date{\today}

\begin{abstract}
We present an \textit{ab initio} quantum theory of the metal-insulator
transition in Ti$_2$O$_3$. The recently developed cluster LDA+DMFT scheme
is applied to describe the many-body features of this compound. 
The conventional single site DMFT cannot reproduce a low temperature
insulating phase for any reasonable values of the Coulomb interaction.
We show that the non-local Coulomb interactions and the strong chemical bonding within Ti-Ti
pair is the origin of the small gap insulating ground state of Ti$_2$O$_3$.
\end{abstract}


\maketitle

The complicated electronic structure and the nature of the metal-insulator
transition in Ti$_{2}$O$_{3}$ and V$_{2}$O$_{3}$ has been the object of
intensive experimental and theoretical investigation over the past half
century \cite{Imada}. Recent progress in high-energy photo-emission
spectroscopy \cite{Spring8} and correlated electrons dynamical-mean field 
theory (DMFT) \cite{review} has shed new light on the first-order metal-insulator
transition in V$_{2}$O$_{3}$. It has been shown that an realistic description of
the metallic and insulating phases of V$_{2}$O$_{3}$ can be obtained from
the combination of a band structure scheme with the local electron-electron
interaction given from DMFT \cite{Held}.
The correlation effects in Ti$_{2}$O$_{3}$ are less clear but angle
resolved photo-emission experiment \cite{Smith} shows a strong reduction of the
Ti 3$d$-bandwidth compared to band structure calculations. The important
question is related to the mechanism of the small, about 0.1 eV, semiconductor
band-gap formation. The generally accepted view is that the metal-insulator
transition is related to the decrease of the $c/a$ ratio in rhombohedral Ti$_{2}$O$_{3}$
and the formation of a Ti-Ti pair along $z$-axis \cite{Goodenough}. Below the
broad (almost 250 K in width) metal-insulator transition at around 470 K the
Ti-Ti pair distance is seen to decrease without any change
of the rhombohedral structure or the formation of long-range antiferromagnetic
order \cite{mag}. This is in contrast to the case of V$_{2}$O$_{3}$ where the V-V pair distance
increases within a monoclinic distortion in the antiferromagnetic phase \cite{Imada}. 

\begin{figure}[tbh]
  \centering
  \begin{minipage}{0.28\textwidth}
     \centering
     \includegraphics[clip=true,width=.5\textwidth]{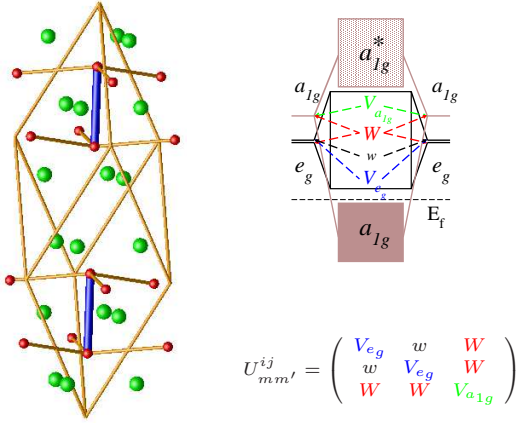}
  \end{minipage}
  \begin{minipage}{0.18\textwidth}
     \begin{flushleft}
        \includegraphics[clip=true,width=.7\textwidth]{scheme.ps}
     \end{flushleft}
     \begin{center}
                                 {\scriptsize
       \begin{equation*}
        \! \! \! \! \! \! \! \! \! \! \! \! \! \! \! U_{mm^{\prime}}^{ij} = \left(
          \begin{array}{ccc}
         \textcolor{blue}{V_{e_g}} &        w                  & \textcolor{red}{W}   \\
                    w              & \textcolor{blue}{V_{e_g}} & \textcolor{red}{W}   \\
         \textcolor{red}{W}        & \textcolor{red}{W}        & \textcolor{green}{V_{a_{1g}}}
          \end{array}
                            \right)
       \end{equation*}           }
    \end{center}
  \end{minipage}
\caption{Left - rhombohedral unit cell of Ti$_{2}$O$_{3}$ corundum
structure. Titanium ions are indicated by the red color, oxygens by green,
and the pair of Ti atoms in $z$ direction by blue. Right -
schematic representation of the $t_{2g}$ splitting in Ti$_{2}$O$_{3}$
(top part) and the intersite Coulomb interaction matrix (bottom part). 
$w$=0 in all our calculations.}
\label{structure}
\end{figure}

Ti$_{2}$O$_{3}$ has an $\alpha $-Al$_{2}$O$_{3}$ corundum structure 
(Fig. \ref{structure}) in the metallic and insulating phases with two formula units
per rhombohedral cell \cite{Abrahams_Rice}. Each Ti atom is surrounded
by the octahedron of oxygens leading to the large $t_{2g}$-$e_{g}^{\sigma }$
splitting. The trigonal distortion gives an additional splitting of $t_{2g}$
bands into $e_{g}^{\pi }$-$a_{1g}$ states and $a_{1g}$ subbands of Ti-Ti pair
form strong bonding-antibonding counterparts (Fig. \ref{structure}). 
In principle, the large decrease of the Ti-Ti distance could split
further an occupied single-degenerate $a_{1g}$ states from a
double-degenerate $e_{g}^{\pi }$ states of $t_{2g}$ subband and form the
insulating $d^{1}$ configuration of this Ti compound. Nevertheless, 
state of the art LDA calculations have shown that for reasonable Ti-Ti pair
distances Ti$_{2}$O$_{3}$ will stay metallic \cite{Mattheiss}.

In order to investigate the role of electron-electron interactions in the
formation of this insulating low-temperature phase one needs an accurate 
estimation of the $a_{1g}$ and $e_{g}$ bandwidths in this complex structure \cite{Zeiger}.
For example a simple free $[$Ti$_{2}$O$_{9}]^{12-}$ cluster mean-field investigation can
easily produce a semiconducting gap due to drastic underestimation of the $a_{1g}$
and $e_{g}$ bandwidths \cite{Nakatsugawa}. On the other hand a more accurate band structure calculation
within the unrestricted Hartree-Fock approximation results in large gap antiferromagnetic state \cite{Catti}.
Thus it is crucial to use both the correct
Green-function embedding of the Ti-Ti pairs as well as a more accurate treatment
of the electron-electron interaction.

The role of metal-metal pair formation and the "molecular" versus band pictures
of the electronic structure have attracted much attention in these compounds \cite{pair}. 
The combination of a strong on-site Coulomb interaction and the large
anisotropy between the hopping parameters in and perpendicular to the pair direction can
favor a localized molecular-orbital picture of the insulating phase. 
However, realistic tight-binding calculations for V$_{2}$O$_{3}$ have shown
the importance of long-range hopping parameters \cite{Elfimov}. It is
also unclear how good an on-site approximation is for the electron-electron
interaction. Since the pair forms a natural "molecular like" element in the corundum-type 
Ti$_{2}$O$_{3}$ structure it might be expected that non-local electron correlations are
important in this system. Thus an approach which combines pair and beyond pair hopping
with non-local electron interactions would be seem to be ideal for this problem.

In this letter we apply for the first time a method, the cluster DMFT scheme \cite{cluster,cdmft}, 
which contains all the physics of correlated pairs in crystals to determine the 
origin of the insulating phase and the metal-insulator transition in Ti$_{2}$O$_{3}$.
A numerically exact multi-orbital Quantum Monte-Carlo (QMC) scheme is deployed for
the solution of the cluster DMFT problem and an accurate first principles tight-binding parametrization
used for the one electronic structure.
Our strategy here is to investigate the gap formation using single site 
\cite{anisimov} and cluster LDA+DMFT with
only local correlations included. We then deploy the full non-local CDMFT and in this way 
are able to directly elucidate the impact non-local Coulomb interactions have on the physics.
We show that the competition between strong bonding within the Ti-Ti pair and
localization from correlation effects leads to the unique situation of the
small semiconducting gap structure in Ti$_{2}$O$_{3}$ oxide and that non-local Coulomb correlations
are of crucial importance for the physics of this small gap insulators.

\begin{figure}[tbp]
\centering
\includegraphics[clip=true,width=.45\textwidth,height=.4\textwidth]{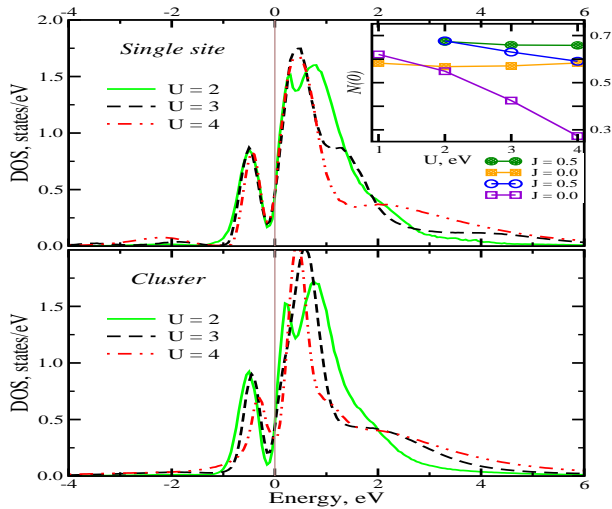}
\caption{Density of states for the single site (upper panel) and cluster
(lower panel) DMFT calculations with different values of the Coulomb repulsion 
$U$ and $J$=0.5 eV. Inset: $N(0)$ $versus$ Coulomb parameter. Filled dark green circles - DMFT
results with $J$=0.5 eV, filled orange squares - DMFT with $J$=0 eV. Open
blue circles and violet squares are CDMFT with $J$=0.5 and 0 eV
respectively.}
\label{dos_ss}
\end{figure}

\begin{figure}[tbp]
\centering
\includegraphics[clip=true,width=.45\textwidth]{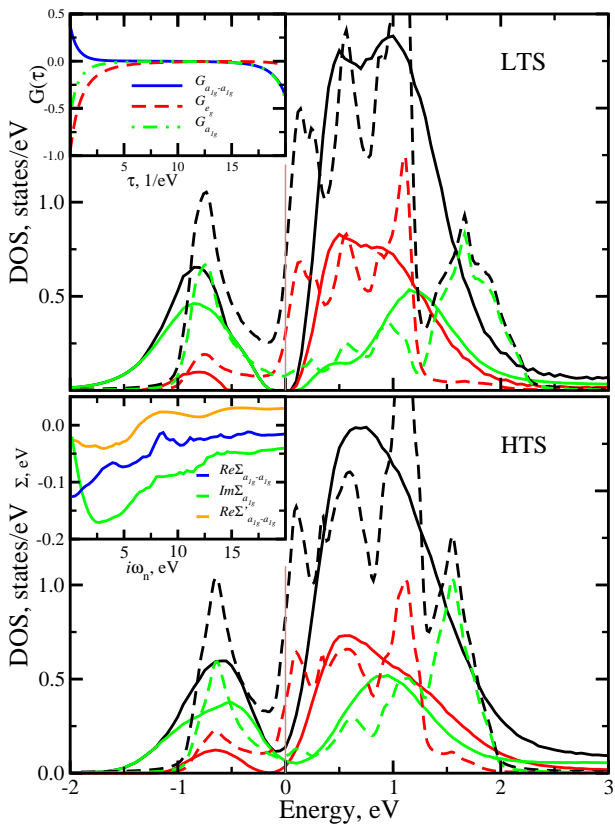}
\caption{Partial and total CDMFT (solid line) compared to the LDA (dashed) DOS 
with $W$=0.5 eV and $V_{a_{1g}}$=$V_{e_{g}}$=0. Total DOS are shown by black,
the $e_{g}$ states by red, and $a_{1g}$ states by green. 
On the upper panel the low temperature structure and $\beta$=20 eV$^{-1}$ are used. 
For the lower panel the high temperature structure and $\beta$=10 eV$^{-1}$ are used.
The diagonal and the biggest $a_{1g}$ off-diagonal Green functions 
$G(\tau)$ are shown in the upper inset. In the lower inset the 
$Re \Sigma_{a_{1g},a_{1g}}$ with $W$=0.5 eV are shown by blue,  
$Re \Sigma^{\prime}_{a_{1g},a_{1g}}$ with $W$=0 eV by orange and
$Im \Sigma_{a_{1g}}$ are shown by green}
\label{dos_2x2}
\end{figure}

We start with the orthogonal LDA Hamiltonian 
$H_{mm^{\prime }}^{LDA}(\mathbf{k})$ 
in the massively downfolded N-th order muffin-tin orbital representation \cite{OKA} 
($m$ corresponds to the 12 $t_{2g}$ orbitals of two Ti-Ti pairs in rhombohedral unit
cell) and include different Coulomb interactions (see Fig. \ref{structure}).
DMFT results for the local and non-local Coulomb interactions are
presented in Figs. \ref{dos_ss},\ref{dos_2x2}.

The bare LDA density of states (DOS) is shown in
Fig. \ref{dos_2x2} by the dashed lines for the low temperature structure (LTS, $\sim$300 K \cite{Abrahams_Rice}) and high temperature 
structural (HTS, $\sim$870 K \cite{Abrahams_Rice}) data on the upper and lower panels respectively. 
Both LTS and HTS electronic structures are metallic within the LDA scheme.
The $a_{1g}$ subband (green dashed line in the Fig. \ref{dos_2x2}) has a strong
bonding-antibonding splitting in contrast to the $e_{g}^{\pi}$ subbands
(red dashed line). The bandwidth of the HTS is approximately 2.8 eV and smaller than
the bandwidth of the LTS (3.2 eV) due to the reduction of 
the $t_{a_{1g},a_{1g}}$ hopping from -0.85 to -0.63 eV. 

The cluster DMFT maps the many-body crystal system onto
an effective self-consistent multi-orbital quantum 
impurity-cluster problem \cite{cluster,cdmft}.
The corresponding Green-function matrix for the Ti-Ti cluster in the LDA+DMFT
scheme is calculated via the BZ-integration
\begin{equation}
\mathbf{G}(i\omega _{n})=\sum_{\mathbf{k}}[(i\omega _{n}+\mu )\mathbf{1}-%
\mathbf{H}^{LDA}(\mathbf{k})-\mathbf{\Sigma }(\omega _{n})]^{-1},  \label{BZI}
\end{equation}
where $\mu$ is the chemical potential defined self-consistently through the
total number of electrons, $\omega _{n}=(2n+1)\pi T$ are the Matsubara
frequencies for temperature $T\equiv \beta ^{-1}$ ($n=0,\pm 1,...$) and 
$\sigma$ is the spin index. The Hamiltonian and the self-energy matrix
have the following super-matrix form corresponding to the
symmetry of two Ti-Ti pairs in the unit cell
\begin{equation}
{\scriptstyle
\left(
\begin{array}{cccc}
  \mathbf{H}_{11}+\mathbf{\Sigma}_{11} & \mathbf{H}_{12}+\mathbf{\Sigma}_{12}
                                       & \mathbf{H}_{13}        & \mathbf{H}_{14} \\
  \mathbf{H}_{21}+\mathbf{\Sigma}_{21} & \mathbf{H}_{22}+\mathbf{\Sigma}_{11} 
                                       & \mathbf{H}_{23}        & \mathbf{H}_{24} \\
  \mathbf{H}_{31}   &   \mathbf{H}_{32}    & \mathbf{H}_{33}+\mathbf{\Sigma}_{11} 
                                       & \mathbf{H}_{34}+\mathbf{\Sigma}_{12}     \\
  \mathbf{H}_{41}   &   \mathbf{H}_{42}    & \mathbf{H}_{43}+\mathbf{\Sigma}_{21} 
                                       & \mathbf{H}_{44}+\mathbf{\Sigma}_{11}
\end{array}
\right), }
\end{equation}   
where $\mathbf{H}_{ij}(\mathbf{k})$ and $\mathbf{\Sigma }_{ij}(\omega _{n})$
are 3$\times $3 matrices for the $t_{2g}$ states and $\mathbf{\Sigma }_{11}$
and $\mathbf{\Sigma }_{12}$ correspond to the intrasite and intersite
contributions to the self-energy respectively.

In the self-consistent cluster
DMFT scheme the local Green-function (\ref{BZI}) should coincide with the
corresponding solution of the effective two-site quantum impurity problem
\cite{review}
\begin{equation}
\mathbf{G}_{\sigma }(\tau -\tau ^{\prime })=-\frac{1}{Z}\int D[\mathbf{c},%
\mathbf{c}^{+}]e^{-S_{eff}}\mathbf{c}(\tau )\mathbf{c}^{+}(\tau ^{\prime })
\label{path}
\end{equation}
here $\mathbf{c}(\tau )$=$[c_{im\sigma }(\tau )]$ is the super-vector of the
Grassman variables, $Z$ is the partition function, $i$ runs over Ti-Ti pair
and $m$ runs over $e_{g}^{\pi }$ or $a_{1g}$ orbitals. The effective
cluster action $S_{eff}$ is defined in terms of so-called ``bath'' Green
function matrix \cite{review}  
$\mathcal{G}_{\sigma}^{-1}(\omega _{n})=\mathbf{G}_{\sigma }^{-1}(\omega _{n})+
\mathbf{\Sigma }_{\sigma }(\omega _{n})$ 
which describes the energy, orbitals, spin and temperature dependent 
interactions of particular cluster with the rest of the crystal 
{\small
\begin{eqnarray}
  S_{eff} & = & -\int_{0}^{\beta} d\tau \int_{0}^{\beta} d\tau^{\prime}
          Tr[\mathbf{c}^{+}(\tau)\mathcal{G}^{-1}(\tau,\tau^{\prime})
	     \mathbf{c}(\tau^{\prime})] +                          \\
     & \frac{1}{2} & \sum_{im,jm^{\prime},\sigma} [ U_{mm^{\prime}}^{ij}
             n_{im}^{\sigma}n_{jm^{\prime}}^{-\sigma}+
	     (U_{mm^{\prime}}^{ij}-J_{mm^{\prime}}^{ij})n_{im}^{\sigma}n_{jm^{\prime}}^{\sigma}]
\notag 
\end{eqnarray}   }
here $n_{im\sigma }$=$c_{im\sigma }^{+}c_{im\sigma }$. We have parameterized 
the screened local Coulomb and exchange matrices ($U_{mm^{\prime }}^{ii}$  and 
$J_{mm^{\prime }}^{ii}$) for the $t_{2g}$ electrons in terms of average
Coulomb and exchange integrals \cite{Kotliar} and used a simple
approximation to the intersite $U_{mm^{\prime }}^{ij}$ interactions as shown in
the Fig. \ref{structure}.

The multi-band impurity QMC scheme \cite{Hirsch,rozenberg} has been used 
for the numerically exact calculation of the cluster Green function (eq. \ref{path}).
The number of auxiliary Ising fields in the discrete Hirsh-Fye transformation were
15 and 58 for the local and non-local Coulomb interaction schemes respectively. For accurate
QMC integration we used
of the order of 10$^{6}$ sweeps, with 8000 $\mathbf{k}$-points for the
BZ-integration. Within 15-20 DMFT iterations convergence
in the self-energy was reached. The maximum entropy method \cite{MEM} has
been used for analytical continuation of the diagonal part of the Green
function matrix to the real energy axis.

Firstly, in Fig.\ref{dos_ss} we show the total density of states 
for both conventional single site (DMFT) and 
the cluster (CDMFT) dynamical mean field theory where 
only local electron correlations have been included.
The QMC simulation has been carried out for $\beta$=20 eV$^{-1}$ which corresponds to a
temperature of $T$$\simeq$580 K which is on the border of the metal-insulator transition.
In the upper panel of Fig. \ref{dos_ss} are shown
the results of DMFT calculations with $U$=2,3,4 eV and exchange parameter $J$=0.5 eV. 
For all values of Coulomb interactions
there is a peak below the Fermi level at around -0.5 eV, predominantly of  $%
a_{1g}$ character with in all cases the same intensity. Above the
Fermi level there are two peaks. The first is at 0.5 eV and has 
$e_{g}$ character while the other peak is strongly dependent on the Coulomb parameter and 
can be associated with an upper Hubbard band. A lower Hubbard band can be seen at around -2 eV. 
We see that for all values of $U$ the the shape of pseudogap is unchanged and
the system remains metallic. 
On the lower part of Fig. \ref{dos_ss} the
results of the CDMFT calculation are shown for the same values of the Coulomb and
exchange parameters. The general structure of the DOS is seen to be similar to the
single site calculation, however one may note interesting differences.
The lower $a_{1g}$ quasiparticle band is decreased in intensity and 
shifted towards the Fermi level from -0.6 eV to -0.3 eV on increasing 
$U$ from 2 to 4 eV. This has the result that for $U$=4 eV the pseudogap
is now located directly at Fermi level, whereas for other 
$U$-values and for all DMFT results it lies on the slope of quasiparticle peak. 

Using the temperature
DOS at the Fermi level, defined as $N(0) \equiv -ImG(\omega_0)/\pi$ with $\omega_0=\pi T$ we are able to 
estimate at what values of $U$ the system will become insulating. This is indicated in the inset in the upper
panel of the Fig. \ref{dos_ss}.
We see that for the single site calculations $N(0)$ depends weakly on $U$ and the
system will remain metallic up to very large values, about 8 eV, of the Coulomb parameter.
On the other hand for the cluster calculation $N(0)$ is seen to decrease
strongly as a function of $U$ for both values of exchange parameter, and the critical
 value for an insulating solution is now lower 
at $U$$\sim$5-6 eV. As expected for the $d^1$ configuration the finite value of the exchange parameter 
effectively decreases the Coulomb interaction matrix. We see the
single site results are in greater contradiction to the experiment as compared 
to LDA (see Fig. \ref{dos_2x2}): 
the local Coulomb interaction leads to the reduction of  
the bonding-antibonding splitting of the $a_{1g}$ subband and this acts to suppress gap formation.
On the other hand in the cluster case a small semiconducting gap
is developed for large $U$ due to dynamical antiferromagnetic correlation in the Ti-Ti pair.

Nevertheless, using either the DMFT or CDMFT schemes with only local correlations there remains
a dramatic absence of gap formation in Ti$_2$O$_3$. We now deploy the full non-local correlation in
CDMFT to the effect of non-local correlations on low and high temperature electronic structure.
We have used different values of the
non-local Coulomb parameters and found that the most important correspond
to non-diagonal interactions. 
For both structures we have chosen values of $U$=2 eV and $J$=0.5 eV 
which are close to those from constrained LDA estimations \cite{uvalue}, while the off diagonal
Coulomb parameter $W$ has been chosen at $W$=0.5 eV \cite{wvalue}.
On the upper panel of Fig. \ref{dos_2x2}  is shown the total and partial DOS for
$\beta$=20 eV$^{-1}$.
Shown also is the LDA result. One can see that for the reasonable parameters chosen
we can reproduce the correct value of the
semiconducting gap $\sim$0.1 eV while keeping the bonding-antibonding splitting on the LDA level.
In the lower panel the high temperature metallic solution corresponding to $\beta$=10 eV$^{-1}$ is shown. 
Here we emphasize that the proper inclusion of the structural effect on the 
LDA level is important as evinced by the fact that
for $\beta$=20 eV$^{-1}$ and high temperature hamiltonian we again obtain a metallic solution.
The $e_g$ states are similar for both LTS and HTS calculations with a small 
shift of occupied part in LTS case. However, the difference 
between the LTS and HTS phases is more pronounced for the $a_{1g}$ states.
The bonding-antibonding splitting in the LTS is about 2 eV whereas 
in the HTS case it is only 1.5 eV. The occupied $a_{1g}$ states in the LTS phase
are shifted down opening the insulating gap. 
The important difference between the large $U$ and small $U$ plus non-local $W$ 
is the absence of well defined Hubbard bands.
This absence makes possible a critical test of the theory proposed here, 
and thus it would be very interesting for photo-emission experiments to check the existence
or not of a lower Hubbard band at around -2 eV.

We have shown that the cluster LDA+DMFT calculation with a moderate
Coulomb repulsion among the $a_{1g}$ orbitals is essential to
produce the high temperature semimetallic state and the low temperature
insulating state.
To understand the role play of the intersite Coulomb interaction
we focus on the the quantity  $t_{a_{1g},a_{1g}} + Re \Sigma_{a_{1g},a_{1g}}(i \omega)$
which we can interpret as a frequency dependent "effective $a_{1g}-a_{1g}$ hopping"
which describes the hopping matrix element in the titanium pair.
We find that this quantity is surprisingly frequency dependent (see lower inset of Fig. \ref{dos_2x2}).

We conclude that  the main role of the intersite Coulomb interaction is dynamic
(the Hartree contribution to this quantity is small) and  results in 
the effective $a_{1g}-a_{1g}$ hopping that {\it increases} as the frequency decreases.  
This enhancement produces a strong level repulsion of the bonding antibonding $a_{1g}$ levels,
lowering the $a_{1g}$ level relative to the $e_g$ level at the low frequency.
This effect combined with a small narrowing of the $a_{1g}$ band opens the $e_g-a_{1g}$
band gap which  results in the insulating state. We checked that this
enhancement of the effective hopping as frequency is decreased is absent
if we turned off the intersite Coulomb repulsion.

This effect is the cluster DMFT analog of a mechanism first discussed in the context
of the single impurity model by Haldane \cite{haldane}. 
He observed, that  a Coulomb repulsion between the impurity level
and additional conduction electron states or screening   channels, {\it enhances} the hybridization
(single impurity analog of the hopping matrix element) as we renormalize to low frequency.

We would like to acknowledge O.K. Andersen, V.I. Anisimov, A. Georges and M.I. Katsnelson 
for useful discussions.
This work was supported by the Netherlands Foundation for the Fundamental Study of Matter (FOM). 
GK was supported by the ONR, grant N000140210766. The authors are grateful to the Kavli Institute of
Theoretical Physics, Santa Barbara, for hospitality during the initial stages of this work.
This research was supported in part by the National Science Foundation under Grant No. PHY99-07949.

\end{document}